\documentclass[twocolumn,  amsmath, amssymb]{revtex4} 
\usepackage{epsfig, amssymb}
\def\be{\begin{eqnarray}}
\def\ee{\end{eqnarray}}
\def\bee{\begin{eqnarray*}}
\def\eee{\end{eqnarray*}}
\newtheorem{thm}{Theorem}

\newtheorem{lem}{Lemma}

\newtheorem{defn}{Definition}

\begin{document}
\title{The invariants of the local Clifford group}
\author{Maarten Van den Nest} \email{mvandenn@esat.kuleuven.ac.be} \author{Jeroen Dehaene}
\author{Bart De Moor} \affiliation{ESAT-SCD, K.U. Leuven, Kasteelpark Arenberg 10, B-3001 Leuven,
Belgium}

\date{\today}
\def\makeheadbox{}

\begin{abstract}
We study the algebra of complex polynomials which remain invariant under the action of the local
Clifford group under conjugation. Within this algebra, we consider the linear spaces of homogeneous
polynomials degree by degree and construct bases for these vector spaces for each degree, thereby
obtaining a generating set of polynomial invariants. Our approach is based on the description of
Clifford operators in terms of linear operations over GF(2). Such a study of polynomial invariants
of the local Clifford group is mainly of importance in quantum coding theory, in particular in the
classification of binary quantum codes. Some applications in entanglement theory and quantum
computing are briefly discussed as well.
\end{abstract}

\maketitle

\section{Introduction}

The (local) Clifford group plays an important role in numerous theoretical investigations, as well
as applications, in quantum information theory, quantum computing and quantum error correction
\cite{QCQI}\cite{loc_per_ent_dist}\cite{entgraphstate}\cite{stab_clif_GF2}\cite{localcliffgraph}\cite{alg_codes}\cite{Gott_heis}.
The Clifford group ${\cal C}_1$ on one qubit consists of all $2\times 2$ unitary operators which
map the Pauli group ${\cal G}_1 = <\sigma_1, \sigma_2, \sigma_3>$ to itself under conjugation,
where $\sigma_1$, $\sigma_2$, $\sigma_3$ are the Pauli matrices. In other words, ${\cal C}_1$ is
the normalizer of ${\cal G}_1$ in the unitary group $U(2)$. The local Clifford group ${\cal C}_n^l$
on $n$ qubits, which is our topic of interest in the following, is the $n$-fold tensor product of
${\cal C}_1$ with itself.

In this paper we study the \emph{invariant algebra} of the local Clifford group, defined as
follows: let $\{\rho_{ij}\}$ be a set of $2^{2n}$ variables, which are assembled in a $2^n\times
2^n$ matrix $\rho=(\rho_{ij})$. The invariant algebra of ${\cal C}_n^l$ then consists of all
complex polynomials $F(\rho)= F(\rho_{11}, \rho_{12}, \dots, \rho_{2^n2^n})$ which remain invariant
under the substitutions $\rho\to U\rho U^{\dagger}$, for every $U\in {\cal C}_n^l$ \footnote{To be
exact, in the literature the invariant algebra of a $N\times N$ matrix group $G$ is usually defined
as the set of all polynomials $p(x)=p(x_1, \dots, x_1)$ such that $p(Ax)=p(x)$ for every $A\in G$.
Our definition is a variant of this.}\label{voetnoot}. It is our goal to construct a generating set
of this algebra.

This research started out as the ground work for the study of equivalence classes of binary quantum
stabilizer codes, the latter being a large and extensively studied class of quantum codes
\cite{Gott}. A stabilizer code is a joint eigenspace of a set of commuting observables in the Pauli
group on $n$ qubits and is described by the projector $\rho_{\cal S}$ on this subspace. Two
stabilizer codes $\rho_{\cal S}$ and $\rho_{{\cal S}'}$ on $n$ qubits are called equivalent if
there exists a local unitary operator $U\in U(2)^{\otimes n}$ such that $U\rho_{{\cal S}}
U^{\dagger}$ is equal to $\rho_{{\cal S}'}$ modulo a permutation of the $n$ qubits. A natural
question to ask is how the equivalence class of a code can be characterized by a minimal set of
invariants, i.e., (polynomial) functions $F(\rho_{\cal S})$ in the entries of the matrix
$\rho_{\cal S}$ which take on equal values for equivalent codes. This is, however, a difficult and
unsolved problem. Therefore, given the explicit connections between stabilizer codes, the Pauli
group and the Clifford group, it seems natural to consider a restricted version of this equivalence
relation, where only operators $U\in{\cal C}_n^l$ are considered, and this is where the invariant
algebra of ${\cal C}_n^l$ comes into play. What is more, it is to date unclear whether this
restriction \emph{is} in fact a restriction at all: indeed, the question exists whether every two
equivalent stabilizer codes are also equivalent in this second, restricted sense. A possible way
towards solving this problem is through a study of invariants (cfr. also \cite{invar_stab}).
Moreover, the problem of recognizing local unitary and/or local Clifford equivalence of certain
classes of multipartite pure quantum states (stabilizer states, graph states) has recently gained
attention both in entanglement theory \cite{entgraphstate}\cite{localcliffgraph}\cite{alg_codes}
and in the one-way quantum computing model \cite{1wayQC}. These examples make for a number of
application domains of the present work.

From a somewhat different perspective, the invariant theory of the Clifford group is also of
interest from a purely mathematical point of view. Runge \cite{runge} and Nebe, Rains and Sloane
\cite{codes_nebe}\cite{cliff_nebe} published a series of papers in which they investigate the
connection between the invariants of the (entire) Clifford group (and generalizations thereof) and
the so-called generalized weight polynomials of a class of self-dual \emph{classical} binary codes.
Their work is a considerable generalization of a central result in classical coding theory, known
as Gleason's theorem \cite{Gleason}, which states that the invariant algebra of ${\cal C}_1$  is
generated by the weight enumerators of the class of doubly-even self-dual classical codes (the
definition of the invariant algebra of ${\cal C}_1$ is here somewhat different than ours, cfr.
footnote 1). It is interesting that the Clifford group - a group which appears naturally in a
quantum theoretical setting, has such a connection, through invariant theory, with the theory of
\emph{classical} codes. It is not known whether this link is a mere coincidence or a manifestation
of some deeper result \cite{rains_self_dual}. This remark may serve as another justification of the
present research.

In our study of the invariant algebra of ${\cal C}_n^l$, we will make extensive use of the
description of this group in terms of binary linear algebra, i.e., algebra over the field
$GF(2)=\Bbb{F}_2$. It is indeed well known that $n$-qubit (local) Clifford operations can be
represented elegantly by a certain class of $2n\times 2n$ linear operators over $\Bbb{F}_2$
\cite{QCQI}\cite{stab_clif_GF2} and this binary picture makes the (local) Clifford group
particularly manageable in the following. In order to obtain a generating set of the invariant
algebra, we will adopt the following basic strategy: note that each invariant polynomial $F$
(simply called \emph{invariant}) can be written as a sum of its homogeneous components, each of
which is an invariant as well. One can therefore always find a generating set of the invariant
algebra which consists of homogeneous invariants only. Furthermore, the set of homogeneous
invariants of fixed degree is a finite-dimensional vector space, as one can easily verify (which
gives the algebra of invariants the structure of a graded algebra). Therefore, a natural approach
to our problem is to consider these spaces of homogeneous invariants degree by degree and to
construct  a basis of invariants for each degree. This construction will yield a generating (yet
infinite) set of the invariant algebra.

\section{The local Clifford group}

The \emph{Clifford group ${\cal C}_1$ on one qubit} is the following group of unitary $2\times 2$
matrices: \[{\cal C}_1 = < \frac{1}{\sqrt 2}\left[
\begin{array}{cc}1 & 1\\1 & -1 \end{array}\right],\  \left[
\begin{array}{cc}1 & 0\\0 & i \end{array}\right]>.\]
The order of ${\cal C}_1$ is finite and equal to 192. Up to overall phase factors, the Clifford
group consists of all unitary operators which map the \emph{Pauli group} to itself under
conjugation; here, the Pauli group ${\cal G}_1$ (on 1 qubit) consists of the identity $\sigma_0$
and the three pauli matrices \[ \sigma_1= \left[
\begin{array}{cc}0 & 1\\1 & 0 \end{array}\right],\ \sigma_2 = \left[
\begin{array}{cc}0 & -i\\i & 0 \end{array}\right],\ \sigma_3=\left[
\begin{array}{cc}1 & 0\\0 & -1 \end{array}\right],\] all having 4 possible overall
phase factors equal to $\pm 1$ or $\pm i$. In other words, up to these overall phase factors, the
group ${\cal C}_1$ is the normalizer of ${\cal G}_1$ in the unitary group $U(2)$. Note that these
phases are not relevant in the following, since we are considering the action of the Clifford group
under conjugation as explained in the introduction. It follows that every $U\in {\cal C}_1$ is, for
our purposes, completely described by a permutation $\pi\in S_3$, where $S_3$ is the symmetric
group on 3 letters, and a set of three phases $\alpha_1$, $\alpha_2$, $\alpha_3=\pm 1$, such that
\[U\sigma_i U^{\dagger} = \alpha_i \sigma_{\pi(i)}\quad (i= 1,2,3).\] Moreover, since
$\sigma_1\sigma_2\sim\sigma_3$, one has $\alpha_{1}\alpha_{2}\alpha_{3}=1$ and it is therefore
sufficient to keep track of only two of the $\alpha_i$'s (say $\alpha_1$ and $\alpha_3$). Another
useful characterization of the Clifford group is obtained by considering the mapping
\begin{eqnarray}\label{map}
\sigma_0=\sigma_{00} &\mapsto& (0,0)\nonumber\\
\sigma_1=\sigma_{01} &\mapsto& (0,1)\nonumber\\
\sigma_3=\sigma_{10} &\mapsto& (1,0)\nonumber\\
\sigma_2=\sigma_{11} &\mapsto& (1,1),
\end{eqnarray}
which establishes a homomorphism between the groups ${\cal G}_1$ and $\Bbb{F}_2^2$. Here,
$\Bbb{F}_2$ is the finite field of two elements (0 and 1), where arithmetics are performed modulo
2. In this representation of Pauli matrices by pairs of bits, a Clifford operation corresponds to
an invertible linear transformation $Q\in GL(2, \Bbb{F}_2)$ (instead of a permutation $\pi\in S_3$)
and a couple of phases $\alpha_1$ and $\alpha_3$. It is this second description of Clifford
operations in terms of binary linear transformations which is most often used in the literature in
quantum information theory and quantum computing, and we will do the same.

The \emph{local Clifford group ${\cal C}_n^l$} on $n$ qubits is the $n$-fold tensor product of
${\cal C}_1$ with itself, i.e. \[{\cal C}_n^l = {\cal C}_1\otimes\dots\otimes{\cal C}_1 \quad (n
\mbox{ times)}.\] Analogous to the case of one qubit, the group ${\cal C}_n^l$ can be most easily
described by its action on the Pauli group ${\cal G}_n$ on $n$ qubits, defined by
\[{\cal G}_n = {\cal G}_1\otimes\dots\otimes{\cal G}_1 \quad (n \mbox{ times)}.\] Using the mapping
(\ref{map}), the elements of ${\cal G}_n$ can be represented as $2n$-dimensional binary vectors as
follows:
\[\sigma_{u_1v_1}\otimes\dots\otimes\sigma_{u_nv_n}=\sigma_{(u,v)} \mapsto  (u,v) \in \Bbb{F}_2^{2n},\]
where $(u,v)=(u_1, \dots, u_n, v_1, \dots,v_n)$. As in the case of one single qubit, local Clifford
operations map ${\cal G}_n$ to itself under conjugation. Therefore, $n$-qubit  local Clifford
operations as well  can be described in terms of linear operations over $\Bbb{F}_2$. One can
readily verify that, in this binary picture, an operator $U\in {\cal C}_n^l$ corresponds to an
invertible $2n\times 2n$ binary matrix $Q$ of the block form
\[Q = \left [ \begin{array}{cc} A&B\\
C&D \end{array}\right],\] where the $n\times n$ matrices $A, B, C, D$ are diagonal, and a set of
$2n$ phases $\alpha_i=\pm 1$, defined by \begin{eqnarray} U\sigma_{e_i}U^{\dagger} =
\alpha_i\sigma_{Qe_i},
\end{eqnarray}
where $e_i$ is the $i$th canonical basis vector in $\Bbb{F}_2^{2n}$, for every $i=1, \dots, 2n$.
Denoting the diagonal entries of $A, B, C, D$, respectively, by
 $a_i$, $b_i$, $c_i$, $d_i$, respectively, the $n$ submatrices \[Q^{(i)}:=\left [
\begin{array}{cc} a_i & b_i
\\ c_i& d_i\end{array} \right ]\in GL(2, \Bbb{F}_2)\] correspond to the tensor factors of $U$.
The group of all such $Q$ is isomorphic to $GL(2, \Bbb{F}_2)^{ n}$ (and $S_3^{n}$).

\section{Invariant polynomials and matrix algebras}

Let $\{\rho_{ij}\}$ be a set of $2^{2n}$ variables, which are assembled in a $2^n\times 2^n$ matrix
$\rho=(\rho_{ij})$. Any homogeneous polynomial $F(\rho)$ of degree $r\in \Bbb{N}_0$ can be written
as a trace
\[F(\rho)=\mbox{Tr }(A_F\cdot \rho^{\otimes r})\] for some complex $2^{nr}\times 2^{nr}$ matrix $A_F$.
To see this, simply note that the tensor product $\rho^{\otimes r}$ contains all monomials of
degree $r$ in the entries $\rho_{ij}$.  The coefficients of these monomials in the polynomial $F$
are encoded in the entries of $A_F$ (note, however, that the correspondence $F \leftrightarrow A_F$
is not one-to-one). It can easily be verified that $F(U\rho U^{\dagger}) = F(\rho)$ for every $U\in
{\cal C}_n^l$ if and only if there exists an $A_F$ such that
\begin{equation}\label{F}
U^{\otimes r}A_F (U^{\otimes r})^{\dagger} =A_F
\end{equation} for every $U\in {\cal C}_n^l$. Therefore, the study of invariant homogeneous
polynomials of fixed degree $r$ is transformed to the study of the algebra ${\cal A}_{n,r}$ of
matrices $A_F$ which satisfy (\ref{F}). In this section, we will construct a linear basis of this
algebra. This will yield a generating set of homogeneous invariants of degree $r$ \footnote{This
set will however not be linearly independent in general, due to fact that the description of an
invariant $F(\rho)$ by a trace $\mbox{Tr }(A_F\cdot \rho^{\otimes r})$ is non-unique. Bases of
invariant polynomials are discussed below.}. First we consider the simplest case of one single
qubit, i.e. $n=1$, and then we move to the general case of arbitrary $n$.

\subsection{One qubit}

Let $r\in \Bbb{N}_0$ and let ${\cal R}_r$ be the averaging operator which maps a $2^r\times 2^r$
matrix  $A$ to
\[{\cal R}_r(A) := \frac{1}{|{\cal C}_1|}\sum_{U\in {\cal C}_1} U^{\otimes r}A (U^{\otimes
r})^{\dagger}.\] Note that ${\cal R}_r$ is the orthogonal projector of the space of $2^r\times 2^r$
matrices onto the subspace ${\cal A}_{1,r}$. Therefore, a spanning (though in general non-minimal)
set of ${\cal A}_{1,r}$ is obtained by fixing a vector space basis of $2^r\times 2^r$ matrices and
calculating its image under ${\cal R}_r$. In this context, a natural choice for such a basis is the
set $\{\sigma_{(u, v)}|\ u,v\in \Bbb{F}_2^r\}$ of Pauli operators on $r$ qubits (all having an
overall phase equal to 1). Before calculating the images ${\cal R}_r(\sigma_{(u,v)})$  in lemma 1,
we need some definitions: firstly, let the group $GL(2, \Bbb{F}_2)$ act on $\Bbb{F}_2^{2r}$ as
follows:
\begin{equation}\label{action} Q\in GL(2, \Bbb{F}_2): (u, v)\in \Bbb{F}_2^{2r} \mapsto (\bar u,
\bar v)\in \Bbb{F}_2^{2r},
\end{equation}
where $(\bar u, \bar v)$ is defined by \[\left[\begin{array}{c}\bar u_j\\
\bar v_j
\end{array}\right]=Q \left[\begin{array}{c} u_j\\ v_j \end{array}\right],\] for every $j=1,
\dots, r$, where $u_j, v_j,\bar u_j,\bar v_j$, respectively, are the components of $u, v,\bar
u,\bar v$.  Secondly, let the binary vector space ${\cal V}_r$ consist of all $(u,
v)\in\Bbb{F}_2^{2r}$ such that
\[\sum_{j=1}^r (u_j,v_j) = (0,0).\]
We are now in a position to state the following lemma:

 \begin{lem} Let $r\in \Bbb{N}_0$. Let
$(u_0,v_0)\in\Bbb{F}_2^{2r}$ and denote by $\Gamma$ the orbit of this vector under the action
(\ref{action}). Then
\[{\cal
R}_r\left(\sigma_{(u_0,v_0)}\right)=
\left\{ \begin{array}{cc}c \sum_{(u, v)\in \Gamma} \sigma_{(u, v)} & \mbox{ if }  (u_0,v_0)\in {\cal V}_r\\
0 &\mbox{ otherwise,}\end{array} \right. \] where $c$ is a constant.
\end{lem}

{\it Proof:} Let $U\in {\cal C}_1$ be an arbitrary Clifford operation. The action of $U$ on the
Pauli matrices is parameterized by coefficients $\alpha_{01}, \alpha_{10}, \alpha_{11}=\pm 1$ with
$\alpha_{01}\alpha_{10}\alpha_{11}=1$ and a linear operator $Q\in GL(2, \Bbb{F}_2)$ such that
$U\sigma_{(a,b)}U^{\dagger}=\alpha_{ab}\sigma_{Q(a,b)}$ for every
$(a,b)\in\Bbb{F}_2^2\setminus\{0\}$. Defining the integers $n_x, n_y, n_z$ by
\begin{eqnarray}
n_x&=&|\{j\ |\ (u_{0j},v_{0j})=(0,1) \}|,\nonumber\\
n_y&=&|\{j\ |\ (u_{0j},v_{0j})=(1,1) \}|,\nonumber\\
n_z&=&|\{j\ |\ (u_{0j},v_{0j})=(1,0) \}|,\nonumber
\end{eqnarray} the operator $U^{\otimes r}$ maps $\sigma_{(u_0,v_0)}$ to
\begin{eqnarray}\label{crucial}
&&\alpha_{01}^{n_x}\alpha_{10}^{n_z}\alpha_{11}^{n_y}\ \sigma_{(\bar u_0,\bar v_0)}=
\alpha_{01}^{n_x+n_y}\alpha_{10}^{n_z + n_y}\ \sigma_{(\bar u_0,\bar v_0)}
\end{eqnarray}
under conjugation, where $(\bar u_0, \bar v_0)\in \Gamma$ is the image of $(u_0,v_0)$ under the
action (\ref{action}) of $Q$. The crucial observation is now that the coefficient of $\sigma_{(\bar
u_0,\bar v_0)}$ in (\ref{crucial}) is always positive (and thus equal to 1) if and only if both the
numbers $n_x+n_y$ and $n_z+n_y $ are even. Note that this occurs if and only if $n_x, n_y $ and
$n_z$ are all even or all odd or, equivalently, if and only if $(u_0,v_0)\in {\cal V}_r$, as one
can readily verify. It follows that \[{\cal R}_r\left(\sigma_{(u_0,v_0)}\right)\sim \sum_{(u, v)\in
\Gamma} \sigma_{(u, v)}\] if $(u_0,v_0)\in {\cal V}_r$. If $(u_0,v_0)\notin {\cal V}_r$, one can
easily see that the different terms in the sum ${\cal R}_r(\sigma_{(u_0,v_0)})$ interfere such as
to yield zero. This ends the proof. \hfill $\square$


Using the result in lemma 1, we can construct a basis of ${\cal A}_{1,r}$. Denote by ${\cal O}_r$
the set of all orbits $\Gamma$ of the elements in ${\cal V}_r$ (note that ${\cal O}_r$ forms a
partition of ${\cal V}_r$). For every $\Gamma\in {\cal O}_r$, define the matrix \be\label{Agamma}
A_{\Gamma}:= \sum_{(u, v)\in \Gamma} \sigma_{(u, v)}.\ee By construction, the matrices $A_{\Gamma}$
linearly generate the algebra ${\cal A}_{1, r}$. Moreover, this set of matrices is linearly
independent: indeed, this follows immediately from the linear independence of the Pauli operators
$\sigma_{(u,v)}$. Therefore, we can conclude that the $A_{\Gamma}$'s are a basis of ${\cal A}_{1,
r}$. In order to calculate the dimension $|{\cal O}_r|$ of ${\cal A}_{1,r}$, we use the
Cauchy-Frobenius orbit-counting lemma, which states that the number of orbits of a finite group $G$
acting on a set $X$ is equal to the average number of fixed points, i.e., the number of orbits is
equal to
\begin{equation}\label{fix}\frac{1}{|G|} \sum_{g\in G} |\mbox{Fix}(g)|,\end{equation} where $|\mbox{Fix}(g)|$ is the number
of fixed points in the set $X$ of the group element $g$. Let us therefore calculate the number of
fixed points of an arbitrary matrix $Q\in GL(2, \Bbb{F}_2)$ acting on ${\cal V}_r$. Firstly, it is
trivial that the identity has $|{\cal V}_r|=4^{r-1}$ fixed points.  Secondly, there are three
elements in $GL(2, \Bbb{F}_2)$ of order two. Consider e.g. the matrix \[ Q_0=\left[
\begin{array}{cc}0 & 1\\1 & 0 \end{array}\right].\] When acting on $\Bbb{F}_2^2$, this operator fixes exactly two
vectors, namely (0,0) and (1,1). Therefore, when $Q_0$ acts on ${\cal V}_r$, the set Fix$(Q_0)$
consists of all vectors of the form \begin{eqnarray} \alpha_1 (1, 0,\dots, 0;\ 1, 0, \dots,0)\ +\
\alpha_2(0, 1,\dots, 0;\ 0, 1, \dots,0)  \nonumber\\+\ \dots\ +\  \alpha_r (0, 0,\dots 1;\ 0, 0,
\dots,1), \mbox{\qquad}
\end{eqnarray}
where $\alpha_i\in \{0,1\}$ for every $i=1, \dots, r$ and where exactly an even number of
$\alpha_i$'s are nonzero. Therefore, the cardinality of $\mbox{Fix}(Q_0)$ is equal to the number of
even subsets of $\{1, \dots, r\}$, i.e. $|\mbox{Fix}(Q_0)|=2^{r-1}$. Note that an analogous
argument holds for the other two matrices of order two. Finally, there are two elements in $GL(2,
\Bbb{F}_2)$ of order three, which fix only the zero vector. Gathering these results in the formula
(\ref{fix}), we find that the number $|{\cal O}_r|$ of orbits  is equal to \[\frac{1}{6}(4^{r-1} +
3\cdot 2^{r-1} + 2).\] We have proven:

\begin{thm}  Let $r\in \Bbb{N}_0$. The set $\{A_{\Gamma}\}_{\Gamma\in {\cal O}_r}$ is a vector space
basis of the algebra ${\cal A}_{1,r}$. The dimension $|{\cal O}_r|$ of ${\cal A}_{1,r}$ is equal to
\begin{equation}\label{dim}\frac{1}{3}(2^{2r-3} + 3\cdot 2^{r-2} + 1).\end{equation}
\end{thm}

Thus, we have obtained the desired result of constructing a basis of matrices of the algebra ${\cal
A}_{1,r}$. It will be useful to have an explicit parameterization of the orbits $\Gamma\in {\cal
O}_r$. Such a parameterization could e.g. be used to enumerate all the matrices $A_{\Gamma}$ for a
given degree. Also when we will move from the matrix algebra ${\cal A}_{1,r}$ to the polynomials
$\mbox{Tr }(A_{\Gamma}\cdot \rho^{\otimes r})$ in section 4, a more operational description of the
$A_{\Gamma}$'s will turn out to be very useful. To this end, for each $(u,v)\in \Bbb{F}_2^{2r}$,
define the sets
\begin{eqnarray}
\eta_{0}(u,v)&=&\{j|\ (u_j, v_j)=(0,0) \},\nonumber\\
\eta_{x}(u,v)&=&\{j|\ (u_j, v_j)=(0,1) \},\nonumber\\
\eta_{y}(u,v)&=&\{j|\ (u_j, v_j)=(1,1) \},\nonumber\\
\eta_{z}(u,v)&=&\{j|\ (u_j, v_j)=(1,0) \}.\nonumber
\end{eqnarray}
Then the following characterization is easily verified: two vectors $(u,v)$, $(u',v')\in
\Bbb{F}_2^{2r}$ belong to the same orbit if and only if
\begin{itemize}
\item[(a)] $\eta_{0}(u,v)=\eta_{0}(u',v')$ and
\item[(b)] there exists a permutation $\pi$ of $\{x, y, z\}$ such that
$\eta_{x}(u',v')=\eta_{\pi(x)}(u,v)$, $\eta_{y}(u',v')= \eta_{\pi(y)}(u,v),$ and $ \eta_{z}(u',v')=
\eta_{\pi(z)}(u,v).$
\end{itemize}
This implies that any orbit $\Gamma$ of the action (\ref{action}) can completely be described by
\begin{itemize}
\item[(a')] a set $\eta_{0}(\Gamma)\subseteq\{1, \dots, r\}$ and
\item[(b')] a partition ${\cal P}(\Gamma)=\{\eta_1, \eta_2, \eta_3\}$
of $\{1, \dots, r\}\setminus\eta_{0}(\Gamma)$ into three (possibly empty) subsets,
\end{itemize}
such that $(u, v)\in \Gamma$ if and only if $\eta_{0}(u,v) = \eta_0(\Gamma)$ and $\{\eta_{x}(u,v),
\eta_{y}(u,v), \eta_{z}(u,v)\}$ $= {\cal P}(\Gamma)$. Moreover, $\Gamma \in {\cal O}_r$ if and only
if the numbers $|\eta_1|, |\eta_2|, |\eta_3|$ are either all even or all odd (cfr. proof of lemma
1).  Let us illustrate this characterization with two simple examples:
\begin{itemize}
\item $r=1$: there is one orbit in ${\cal O}_1$, namely $\Gamma_0=\{(0,0)\}\in {\cal O}_1.$ This orbit is characterized by
$\eta_{0}(\Gamma_0) = \{1\}$ and ${\cal P}(\Gamma_0) = \{\emptyset, \emptyset, \emptyset\}$.
\item $r=2$: there are two orbits in ${\cal O}_2$,
namely $\Gamma=\{(0,0;0,0)\}$ and \begin{eqnarray}\Gamma'&=&\{(0,0;1,1), (1,1;0,0), (1,1;1,1)\}\nonumber\\
&=& \{(u,v)\in \Bbb{F}_2^4| (u_1, v_1) = (u_2, v_2) \neq (0,0)\}.\nonumber\end{eqnarray} The orbits
$\Gamma$ and $\Gamma'$  are described by \[\eta_{0}(\Gamma) = \{1,2\},\ {\cal P}(\Gamma) =
\{\emptyset, \emptyset, \emptyset\}\] and \[\eta_{0}(\Gamma') = \emptyset,\ {\cal P}(\Gamma') =
\left\{ \{1,2\}, \emptyset, \emptyset\right\}.\]
\end{itemize}

\subsection{Multiple qubits}

For arbitrary $n$, the result in theorem 1 can immediately be used to construct a basis of ${\cal
A}_{n,r}$. To see this, let us first consider the algebra of $2^{nr}\times 2^{nr}$ matrices $A$
which satisfy \[U_1^{\otimes r}\otimes\dots\otimes U_n^{\otimes r} A (U_1^{\otimes
r}\otimes\dots\otimes U_n^{\otimes r})^{\dagger} = A,\] for every $U_1, \dots, U_n\in {\cal C}_1$.
It is straightforward to show that this algebra is the $n$-fold tensor product of ${\cal A}_{1,r}$
with itself. Therefore, a basis of this algebra is given by the matrices
$A_{\Gamma_1}\otimes\dots\otimes A_{\Gamma_n}$, where $\Gamma_i$ ranges over all orbits in ${\cal O
}_r$, for every $i=1, \dots, n$. In order to obtain a basis of ${\cal A}_{n, r}$, one simply has to
conjugate this basis with the permutation matrix $P$, defined by
\begin{eqnarray}&&P\ | i_{11}\dots i_{1r}; i_{21}\dots i_{2r};\dots; i_{n1}\dots i_{nr}\rangle
\nonumber\\ & & = | i_{11}\dots i_{n1}; i_{12}\dots i_{n2};\dots; i_{1r}\dots
i_{nr}\rangle,\end{eqnarray} where $i_{ab}\in \{0,1\}$ and $|i_{11}\dots\rangle$ are the standard
basis vectors in $\Bbb{C}^{2^{nr}}$. Indeed, the matrix $P$ performs the appropriate permutation of
tensor factors, mapping $U_1^{\otimes r}\otimes\dots\otimes U_n^{\otimes r}$ to
$(U_1\otimes\dots\otimes U_n)^{\otimes r}$ under conjugation. This leads to the following result:

\begin{thm}  Let $r\in \Bbb{N}$. For every $n$-tuple $\gamma=(\Gamma_1, \dots, \Gamma_n)$ of orbits
$\Gamma_i\in {\cal O}_r$, define the matrix
\begin{eqnarray}\label{Agammatuple}A_{\gamma}&:=& P A_{\Gamma_1}\otimes\dots\otimes A_{\Gamma_n}P^T.\end{eqnarray}
Then the set $\{ A_{\gamma}\}_{\gamma}$ forms a vector space basis of ${\cal A}_{n, r}$. The
dimension of ${\cal A}_{n, r}$ is equal to $|{\cal O}_r|^n$.
\end{thm}

Following the discussion at the end of section 3.1., the matrices $A_{\gamma}$ can be described in
an alternative way than (\ref{Agammatuple}), using the description of orbits $\Gamma\in{\cal O}_r$
by couples $(\eta_{0}(\Gamma), {\cal P}(\Gamma))$. Defining the \emph{support} of a vector $w\in
\Bbb{F}_2^{2n}$ to be the set \be\label{supp} \mbox{supp}(w) = \{i\in \{1, \dots, n\}\ |\
(w_i,w_{n+i})\neq(0,0)\},\ee one obtains:

\begin{thm}Let $\gamma=(\Gamma_1, \dots, \Gamma_n)$ be an $n$-tuple of orbits $\Gamma_i\in {\cal O}_r$.
For every $j, k\in\{1,\dots, r\}$, $j<k$, define the sets $\omega^{(j)}$ and $\omega^{(jk)}$ by
\begin{eqnarray}
&\omega^{(j)}&= \{i\in \{1, \dots, n\}\ |\ j\in \eta_0(\Gamma_i)\}\label{omegas1}\nonumber\\
&\omega^{(jk)} &= \{i\in \{1, \dots, n\}\ |\ j, k \in \eta_0(\Gamma_i)\mbox{ or j and k}\nonumber\\
&&\mbox{\quad belong to the same  subset of } {\cal P}(\Gamma_i) \}.\label{omegas2}
\end{eqnarray} Then
$A_{\gamma}=\sum \ \sigma_{w^{(1)}} \otimes\dots\otimes \sigma_{w^{(r)}},$ where the sum runs over
all ordered $r$-tuples $(w^{(1)},\dots, w^{(r)})\in (\Bbb{F}_2^{2n})^{\times r}$ satisfying
\begin{eqnarray}
&&\mbox{ supp}(w^{(j)})= \bar \omega^{(j)} \label{supps1}\\
&&\mbox{\ supp}(w^{(j)}+ w^{(k)})= \bar \omega^{(jk)},\label{supps2}
\end{eqnarray}
for every $j, k\in\{1,\dots, r\}$, $j<k$, where $\bar \omega^{(j)}$, $\bar \omega^{(jk)}$ denote
the complements of the sets $ \omega^{(j)}$, $ \omega^{(jk)}$ in $\{1, \dots, n\}$.
\end{thm}

{\it Proof:} By definition, $A_{\gamma}$ is equal to
\[\sum\ \sigma_{w^{(1)}} \otimes\dots\otimes \sigma_{w^{(r)}},\] where the sum runs over all
ordered $r$-tuples $(w^{(1)},\dots, w^{(r)})\in (\Bbb{F}_2^{2n})^{\times r}$ such that
\be(w_i^{(1)},\dots, w_i^{(r)},w_{n+i}^{(1)},\dots, w_{n+i}^{(r)})\in \Gamma_i,\ee for every $i=1,
\dots,n$. The proof of the theorem then follows immediately from the characterization of the orbits
$\Gamma_i$ by the couples $(\eta_{0}(\Gamma_i), {\cal P}(\Gamma_i))$, for every $i=1, \dots, n$.
\hfill $\square$

{\it Example 1.} Let us consider this result for the case of smallest nontrivial degree, i.e.
$r=2$. Let $\gamma^{(2)}=(\Gamma_1, \dots, \Gamma_n)$ be an $n$-tuple  of orbits $\Gamma_i\in {\cal
O}_2$. Recall that ${\cal O}_2$ contains exactly two orbits $\Gamma$ and $\Gamma'$, as defined in
the last paragraph of section 3.1. Let $\omega$ be the subset of $\{1, \dots, n\}$ which consists
of all $i$ such that $\Gamma_i = \Gamma$. Following the definitions stated in theorem 3, we have
$\omega^{(1)} = \omega= \omega^{(2)}$ and $\omega^{(12)} = \{1, \dots, n\}$. Consequently
\[A_{\gamma^{(2)}} = \sum_{w \in \Bbb{F}_2^{2n}, \mbox{\scriptsize\ supp}(w)= \bar\omega } \sigma_w
\otimes \sigma_w.\] This shows that the matrices $A_{\gamma^{(2)}}$ are parameterized by the
subsets $\omega$ of $\{1, \dots,n\}$ in a one-to-one correspondence.

While the result in theorem 3 is in fact no more than a reformulation of (\ref{Agammatuple}), it is
interesting in that it relates the matrices $A_{\gamma}$ (and thus the corresponding invariant
polynomials $\mbox{Tr }(A_{\gamma}\cdot \rho^{\otimes r})$ as well) to the notion of the support of
a binary vector, which is of central importance in quantum coding theory. Note that the definition
(\ref{supp}) of support is indeed the same as is used in the theory of quantum codes.

\section{Bases of invariants}

It  follows from theorem 2 that the polynomials
\begin{equation}\label{pol1bit}p_{n,r}^{\gamma}(\rho):=\mbox{Tr }(A_{\gamma}\cdot \rho^{\otimes
r}),\end{equation} in the variables $\rho_{ij}$ ($i, j = 0,\dots, 2^n -1$) linearly generate the
space of homogeneous invariants of ${\cal C}_n^l$ of degree $r$. However, different $A_{\gamma}$'s
may correspond to the same polynomial and therefore linear dependencies within the set of the
polynomials (\ref{pol1bit}) can exist in general. We now set out to pinpoint a basis of polynomials
for each degree $r$. As in the preceding section, we start by considering the simplest case of one
qubit and then move to the general case.

\subsection{One qubit}

Let $\rho=(\rho_{ij})$, where $i,j=0,1$, be a matrix of variables. Fix an orbit $\Gamma \in {\cal
O}_r$ with $\eta_0(\Gamma)\equiv \eta_0$ and ${\cal P}(\Gamma) \equiv \{\eta_1, \eta_2, \eta_3\}$.
It will be convenient to introduce the linear forms $x_{ij}(\rho):= \mbox{ Tr}(\rho \sigma_{ij})$,
where $i, j=0, 1$,  or more explicitly:
\begin{eqnarray}\label{x_rho}
x_{00}(\rho) &=& \rho_{00} + \rho_{11}\nonumber\\
x_{01}(\rho) &=& \rho_{01} + \rho_{10}\nonumber\\
x_{10}(\rho) &=& \rho_{00} - \rho_{11}\nonumber\\
x_{11}(\rho) &=& i(\rho_{01} - \rho_{10}).
\end{eqnarray}
Conversely,  the $\rho_{ij}$'s can be written as  linear forms in the variables $x=(x_{ij})$ as
follows:
\[\rho(x)= \frac{1}{2} \sum_{i,j=0}^1 x_{ij}\sigma_{ij}.\] We will consider $\mbox{Tr
}(A_{\Gamma}\cdot \rho(x)^{\otimes r})$ to be a polynomial in the variables $x$. This yields
\begin{eqnarray}
\mbox{Tr }(A_{\Gamma}\cdot \rho(x)^{\otimes r}) = \frac{1}{2^r}\sum_{(u,v)\in \Gamma}
x_{u_1v_1}\dots
x_{u_rv_r}\nonumber\\
=\frac{1}{2^r}\sum_{(u,v)\in \Gamma} x_{00}^{n_0(u, v)}x_{01}^{n_x(u, v)}x_{10}^{n_z(u,
v)}x_{11}^{n_y(u, v)},\label{pol1bit2}
\end{eqnarray}
where we have used the definitions $n_0(u, v) = |\eta_0(u, v)|$ etc.. Note that \[n_0(u, v) =
|\eta_0|\] and \[\{n_x(u, v), n_y(u, v), n_z(u, v)\}=\{|\eta_1|, |\eta_2|, |\eta_3|\}\] for every
$(u,v)\in \Gamma$. It readily follows that (\ref{pol1bit2}) is equal to
\begin{equation}\label{pol1bit3}
x_{00}^{|\eta_0|} \sum_{\pi\in S_3} x_{01}^{|\eta_{\pi(1)}|} x_{10}^{|\eta_{\pi(2)}|}
x_{11}^{|\eta_{\pi(3)}|}
\end{equation}
up to a normalization factor.
Expression (\ref{pol1bit3}) shows that the polynomial $\mbox{Tr }(A_{\Gamma}\cdot \rho^{\otimes
r})$ only depends on the number $|\eta_0|$ and the set $\{|\eta_1|, |\eta_2|, |\eta_3|\}$. In other
words, if $\Gamma$ and $\Gamma'$ are two orbits such that  \[|\eta_0(\Gamma)| = |\eta_0(\Gamma')|\]
and
\[\{|\eta_1(\Gamma)|, |\eta_2(\Gamma)|, |\eta_3(\Gamma)|\}=
\{|\eta_1(\Gamma')|, |\eta_2(\Gamma')|, |\eta_3(\Gamma')|\},\] then (and only then) the polynomials
$\mbox{Tr }(A_{\Gamma}\cdot \rho^{\otimes r})$ and $\mbox{Tr }(A_{\Gamma'}\cdot \rho^{\otimes r})$
coincide. This equivalence relation on ${\cal O}_r$ leads to the following definition: for each
4-tuple $\lambda=(\lambda_0, \lambda_1, \lambda_2, \lambda_3)$ of non-negative integers $\lambda_i$
such that $\lambda_1, \lambda_2,\lambda_3$ are either all even or all odd, $\lambda_0 + \lambda_1+
\lambda_2+ \lambda_3= r$ and $\lambda_1\geq \lambda_2\geq \lambda_3$, we define an invariant
$p_{r}^{\lambda}$ of ${\cal C}_1$ of degree $r$ as follows:
\begin{equation}\label{pol1bit4}
p_{r}^{\lambda} = x_{00}^{\lambda_0} \sum_{\pi\in S_3} x_{01}^{\lambda_{\pi(1)}}
x_{10}^{\lambda_{\pi(2)}} x_{11}^{\lambda_{\pi(3)}}.
\end{equation}
Recall that $p_{r}^{\lambda}$ is to be regarded as a polynomial in the variables $\rho$ via
(\ref{x_rho}). By construction, the set of all these polynomials generates the space of invariants
of degree $r$. What is more, the $p_{r}^{\lambda}$'s are linearly independent. This immediately
follows from the fact that each monomial in the variables $x_{ij}$ occurs in exactly one polynomial
$p_{r}^{\lambda}$ and that the polynomials $x_{ij}(\rho)$ are algebraically independent. We have
therefore proven:

\begin{thm} The polynomials $p_{r}^{\lambda}$ form a basis of the vector space of homogeneous
invariants of ${\cal C}_1$ of degree $r$.
\end{thm}


\subsection{Multiple qubits}

The construction of bases of invariants for arbitrary $n$ will be a generalization of the one qubit
case. Starting from a $2^n\times 2^n$ matrix $\rho$ of variables, we again perform a change of
variables, defining $x_{w}\equiv x_w(\rho)=\mbox{ Tr}(\rho\cdot \sigma_{w})$, for every $w\in
\Bbb{F}_2^{2n}$. Analogous to the one qubit case, the converse relation reads
$\rho(x)=\frac{1}{2^n} \sum_w x_w\sigma_w$. Note that the polynomials $\{x_{w}(\rho)\}$ are
algebraically independent; this follows from the fact that the variables $x$ and the variables
$\rho$ are related by an invertible linear transformation. Now, letting $\gamma=(\Gamma_1, \dots,
\Gamma_n)$ be an $n$-tuple of orbits $\Gamma_i\in{\cal O}_r$, the invariant $p_{n,r}^{\gamma}$,
regarded as a polynomial in the variables $x$, is equal to \be\label{polmult} \sum_{(w^{(1)},
\dots, w^{(r)})\in \gamma} x_{w^{(1)}}\dots x_{w^{(r)}} \ee up to a normalization. Here, $(w^{(1)},
\dots, w^{(r)})\in \gamma$ is a shorthand notation to express that $(w^{(1)}, \dots, w^{(r)})$ is
an $r$-tuple of vectors $w^{(j)}\in\Bbb{F}_2^{2n}$ satisfying \be(w_i^{(1)},\dots,
w_i^{(r)},w_{n+i}^{(1)},\dots, w_{n+i}^{(r)})\in \Gamma_i,\ee for every $i=1, \dots, n$. As in the
case of one single qubit, the correspondence between the polynomial $p_{n,r}^{\gamma}$ and the
matrix $A_{\gamma}$ is non-unique. Indeed, suppose that $\mu\in S_r$ is an arbitrary permutation
and define the $n$-tuple $\gamma^{\mu}=(\Gamma_1^{\mu}, \dots, \Gamma_n^{\mu})$ such that \be
j\in\eta_a(\Gamma_i^{\mu}) \mbox{\quad iff\quad } \mu^{-1}(j)\in\eta_a(\Gamma_i)  \ee for every
$j\in\{1,\dots, r\}$ and $a\in\{0,1,2,3\}$. Equivalently, one has $(w^{(1)}, \dots, w^{(r)})\in
\gamma^{\mu}$ if and only if $(w^{(\mu(1))}, \dots, w^{(\mu(r))})\in \gamma$. Then
\be\label{perm_sym}p_{n,r}^{\gamma}=p_{n,r}^{\gamma^{\mu}},\ee which immediately follows from
(\ref{polmult}). Conversely, if $\gamma$ and $\gamma'$ are two $n$-tuples of orbits such that
$p_{n,r}^{\gamma}=p_{n,r}^{\gamma'}$, then there exists a permutation $\mu\in S_r$ such that
$\gamma' = \gamma^{\mu}$, as one can easily verify. We now claim that a basis
$\{p_{n,r}^{\gamma_1}, p_{n,r}^{\gamma_2},\dots\}$ of the space of invariants of ${\cal C}_n^r$ is
obtained by fixing a set $\{\gamma_1, \gamma_2,\dots\}$ of $n$-tuples of orbits such that
\begin{itemize}
\item[(i)] The polynomials $p_{n,r}^{\gamma_i}$ are pairwise different
\item[(ii)] For every $n$-tuple $\gamma$ of orbits, $p_{n,r}^{\gamma}=p_{n,r}^{\gamma_i}$ for some $i=1, 2,\dots$.
\end{itemize}
The claim is proven as follows: firstly, it follows from the construction of the invariants
$p_{n,r}^{\gamma}$ and item (ii) that the polynomials $p_{n,r}^{\gamma_i}$ generate the space of
homogeneous invariants of degree $r$. Secondly, the linear independence of the
$p_{n,r}^{\gamma_i}$'s follows from (i). For, suppose there exist complex coefficients $a_i$, not
all equal to zero, such that \be \sum_i a_ip_{n,r}^{\gamma_i}=0. \ee As each monomial
$\prod_{j=1}^r x_{w^{(j)}}$, where $w^{(j)}\in \Bbb{F}_2^{2n}$, occurs in exactly one invariant
$p_{n,r}^{\gamma_i}$, this yields a nontrivial linear combination of these monomials adding up to
zero, which is a contradiction; indeed, the monomials $\prod_{j=1}^r x_{w^{(j)}}$ are linearly
independent, as the polynomials $\{x_{w}(\rho)\}$ are algebraically independent.

We now set out to construct a set of invariants which satisfies (i)-(ii). According to the
discussion above, there is an equivalence relation $\sim$ on the set ${\cal O}_r^n$ of $n$-tuples
of orbits, such that $\gamma\sim\gamma'$ if and only if there exists a permutation $\mu\in S_r$
such that $\gamma'=\gamma^{\mu}$. A set of invariants which satisfies the desired conditions is
obtained by choosing any set $\{\gamma_1, \gamma_2, \dots\}$ of orbits such that every equivalence
class is represented by exactly one $n$-tuple $\gamma_i$.


Recall that an $n$-tuple $\gamma=(\Gamma_1, \dots, \Gamma_n)\in{\cal O}_r^n$ is described by $n$
couples $(\eta_0(\Gamma_i),{\cal P}(\Gamma_i))$, where $\eta_0(\Gamma_i)\subseteq\{1, \dots, r\}$
and ${\cal P}(\Gamma_i)$ is a partition of $\{1, \dots, r\}\setminus\eta_0(\Gamma_i)$ into three
subsets. While such a system of $n$ couples compactly describes $\gamma$, it will be useful to
represent $\gamma$ in a different way, which contains some redundant information but has the
advantage of being more transparent: we describe $\gamma$ by an $n\times r$ matrix $M$ with entries
in the set $\{0,1,2,3\}$, satisfying \be\label{M} &&M_{ij} = 0 \mbox{\quad iff } j\in
\eta_0(\Gamma_i),\nonumber\\ &&0\neq M_{ij}=M_{ik} \mbox{ iff j and k belong to the same
subset}\nonumber\\&&\mbox{\qquad\qquad\qquad\qquad in the partition } {\cal P}(\Gamma_i), \ee for
every $i=1, \dots, n$ and $j,k=1, \dots, r$. It is clear that this description exhibits some
degeneracy, as any permutation of $\{1,2,3\}$ in any row of $M$ yields a (generally) different
matrix which also satisfies (\ref{M}). However, the equivalence relation $\sim$ is translated into
a simple kind of equivalence transformation of matrices. Indeed, two $n$-tuples
$\gamma,\gamma'\in{\cal O}_r^n$, described by $n\times r$ matrices $M$ and $M'$, respectively,
belong to the same equivalence class of the relation $\sim$ if and only if $M'$ is equal to $M$
modulo a permutation $\mu\in S_r$ of its columns and $n$ row-wise permutations $\pi_i$ of $\{1, 2,
3\}$, and we write $M\sim M'$.

Seeing that we are looking for suitable representatives of each equivalence class, it is
appropriate to look for normal forms of the matrices $M$ under the above action of the permutations
$\mu$ and $\pi_i$. There is in fact a lot of freedom to define sensible normal forms. One possible
definition is stated below in definition 4. First we need some preliminary definitions:
\begin{defn}
Let $d\in\Bbb{N}_0$. Let $u = (u_1, u_2, \dots, u_d)$, $v = (v_1, v_2, \dots, v_d)$ be two
$d$-dimensional vectors with nonnegative integer components. A lexicographical ordering relation
$\leq_{\mbox{\scriptsize} lex}$ is defined as follows: $u \leq_{\mbox{\scriptsize lex}} v$ if $u=v$
or if there exists $j$ ($1 \leq j \leq d$) such that $u_i=v_i$ if $i < j$ and $u_j<v_j$.
\end{defn}

\begin{defn}
Let $u$ be a $d$-dimensional vector with entries in $\{0,1,2,3\}$. For every $a\in\{0,1,2,3\}$,
define $\eta_{a}(u)=\{j\in\{1,\dots, d\}\ |\ u_j= a\}$.
\end{defn}

\begin{defn}
Let $M$ be an $n\times r$ matrix with entries in the set $\{0,1,2,3\}$. Let $M_i^T$ denote the
$i$th row of $M$. Let $m=(m_1, \dots, m_{i_0})$ be an $i_0$-dimensional vector with entries in
$\{0,1,2,3\}$, where $i_0\leq n$. Then the set $\eta_m(M)\subseteq\{1, \dots r\}$ is defined as
follows: \be \eta_m(M) = \cap_{i\leq i_0} \eta_{m_i}(M_i^T).\ee For every $a\in\{1,2,3\}$, the
vector $u_{i_0+1}^{(a)}(M)$ with components $u_{i_0+1}^{(a)}(M)_{m}$, where $m$ ranges over all
$i_0$-dimensional vectors with components in $\{0,1,2,3\}$, is defined by \be\label{u}
u_{i_0+1}^{(a)}(M)_{m} = |\{j\in \eta_m(M)| M_{i_0+1,j}= a\}| \ee (the indices $m$ of the
components of $u_{i_0+1}^{(a)}(M)$ are ordered according to the lexicographical ordering relation.)
\end{defn}

\begin{defn}
Let $M$ be an $n\times r$ matrix with entries in the set $\{0,1,2,3\}$. Then $M$ is in normal form
if it satisfies the following conditions:
\begin{itemize}
\item[(i)] The columns $K_j$ of $M$ are ordered non-decreasingly, i.e.
$K_1\leq_{\mbox{\scriptsize lex}}\dots\leq_{\mbox{\scriptsize lex}}K_r$
\item[(ii)] $|\eta_{3}(M_1^T)|\leq|\eta_{2}(M_1^T)|\leq|\eta_{1}(M_1^T)|$ and for every $i=2, \dots, n$,
 \be u_{i}^{(3)}(M)\leq_{\mbox{\scriptsize lex}}
u_{i}^{(2)}(M)\leq_{\mbox{\scriptsize lex}}u_{i}^{(1)}(M). \ee
\item[(iii)] For every $i=1, \dots, n$ the three numbers $|\eta_1(M_i^T),\ |\eta_2(M_i^T)|,\ |\eta_3(M_i^T)|$
are either all even or all odd.
\end{itemize}
\end{defn}

{\it Example 2.} The following $3\times 11$ array is in normal form: \be \left[
\begin{array}{c}0\ 0\ 0\quad 111\ 1\quad 22\quad 33\\0\ 1\ 2\quad 111\
2\quad 33\quad 22\\1\ 2\ 3\quad 012\ 3\quad 03\quad 12
 \end{array} \right].\ee Indeed, conditions (i) and and (iii) are easily checked,
as well as the first part of condition (ii). As for the second part of (ii), let us calculate the
vectors \be u_2^{(a)}(M)= \left( (u_2^{(a)})_0, (u_2^{(a)})_1, (u_2^{(a)})_2,
(u_2^{(a)})_3\right)\ee and $u_3^{(a)}(M)$ is equal to \be \left((u_3^{(a)})_{00},
(u_3^{(a)})_{01}, (u_3^{(a)})_{02}, (u_3^{(a)})_{03},(u_3^{(a)})_{10}, (u_3^{(a)})_{11},\dots
\right).\nonumber\ee Using definition (\ref{u}), we find \be u_2^{(1)}= (1,3,*,*),\ u_2^{(2)}=
(1,1,*,*),\
u_2^{(3)}= (0,0,*,*)\nonumber \ee and \be u_3^{(1)}&=& (1,0,0,*,\dots)\nonumber\\ u_3^{(2)}&=& (0,1,0,*,\dots)\nonumber\\
u_3^{(3)}&=& (0,0,1,*,\dots),\ee where the entries denoted with $*$ are (in this example)
irrelevant to order the vectors lexicographically, and condition (ii) follows. \hfill $\diamond$

One can easily verify that each equivalence class contains exactly one normal form. Note that,
given an $n\times r$ normal form $M$, one recovers the corresponding tuple $\gamma_M=(\Gamma_1,
\dots, \Gamma_n)\in{\cal O}_r^n$ as follows: \be \eta_0(\Gamma_i)&=&\eta_0(M_i^T) \nonumber\\ {\cal
P}(\Gamma_i) &=& \left\{\eta_1(M_i^T), \eta_2(M_i^T), \eta_3(M_i^T)\right\}.\ee For instance, the
tuple $\gamma$ corresponding to the normal form in example 2 is defined by: \be
\eta_0(\Gamma_1)&=&\{1,2,3\},\ {\cal P}(\Gamma_1) = \{\{4,5,6,7\}, \{8,9\},
\{10,11\}\},\nonumber\\
\eta_0(\Gamma_2)&=&\{1\},\ {\cal P}(\Gamma_2) = \{\{2,4,5,6\}, \{3,7,10,11\}, \{8,9\}\},\nonumber\\
\eta_0(\Gamma_3)&=&\{4,8\},\ {\cal P}(\Gamma_3) = \{\{1,5,10\}, \{2,6,11\}, \{3,7,9\}\}.\nonumber
\ee We have proven our main result:
\begin{thm}  For every  $n\times r$ normal form $M$, denote the corresponding $n$-tuple of orbits by $\gamma_M$.
Then the set of all invariants $p^{\gamma_M}_{n,r}$ forms a basis of the space of homogeneous
invariants of ${\cal C}_n^l$ of degree $r$.
\end{thm}
Thus, we have obtained our initial objective of constructing for every $n$ and for every $r$ a
basis of the space of invariants of ${\cal C}_n^l$ of degree $r$. Note that for the case $n=1$ we
indeed recover the result obtained in the previous section.

It is interesting to investigate the behavior of the dimensions $d_{n,r}$ of these spaces for large
$n$ and $r$. Lower and upper bounds for $d_{n,r}$ are the following:
\begin{lem}  Let $n,r\in\Bbb{N}_0$. Then \be \frac{1}{6^nr!}(4^{r-1} + 3\cdot 2^{r-1} + 2)^n
\leq d_{n,r}\leq \left( \begin{array}{c} r + 4^n-1\\ r\end{array}\right)\nonumber\ee
\end{lem}

{\it Proof:} Let ${\cal M}_{n\times r}$ denote the set of all $n\times r$ matrices $M$ with entries
in the set $\{0,1,2,3\}$, such that for every $i=1, \dots, n$ the three numbers \be|\eta_1(M_i^T),\
|\eta_2(M_i^T)|,\ |\eta_3(M_i^T)|\ee are either all even or all odd. Recall that $d_{n,r}$ is equal
to the number of orbits of the group $S_r\times S_3^n$ acting on this set as defined above. Using
the Cauchy-Frobenius lemma, the number of orbits is equal to \be
\label{upperlowerbnd}\frac{1}{6^nr!} \sum_{(\mu,\pi_i)} \mbox{ Fix}(\mu,\pi_i),\ee where $\mbox{
Fix}(\mu,\pi_i)$ denotes the number of fixed points in ${\cal M}_{n\times r}$ of the element
$(\mu,\pi_i)=(\mu, \pi_1, \dots, \pi_n)$, where $\mu\in S_r$ and $\pi_i\in S_3$. Firstly, note that
restricting the sum to all group elements were $\mu$ is equal to the identity yields the desired
lower bound, using a highly similar argument to the one used to calculate $|{O}_r|^n$ above. In
order to obtain the upper bound, we will calculate the number $N_{n,r}$ of orbits of the group
$S_r$ acting on the set of \emph{all} $n\times r$ matrices with entries in the set $\{0,1,2,3\}$ by
permuting columns. Note that this number is indeed an upper bound for $d_{n,r}$. The
Cauchy-Frobenius lemma yields \be N_{n,r} =\frac{1}{r!}\sum_{\mu\in S_r} (4^n)^{c(\mu)},\ee where
$c(\mu)$ denotes the number of cycles in the permutation $\mu$. Consequently \be N_{n,r}
=\frac{1}{r!}\sum_{k=0}^r t(r,k)4^{nk},\ee where $t(r,k)$ is defined as the number of permutations
in $S_r$ which have exactly $k$ cycles. Note that this number is related to the \emph{Stirling
number $s(r,k)$ of the first kind} by the relation $t(r,k) = (-1)^{r+k}s(r,k)$ \cite{stirling}.
Using the
identity \cite{stirling} \be \sum_{k=0}^r  s(r,k)x^k= (-1)^r r! \left( \begin{array}{c} r -x-1\\
r\end{array}\right),\ee we find that \be N_{n,r} = \left( \begin{array}{c} r +4^n-1\\
r\end{array}\right),\ee which completes the proof. \hfill$\square$

While these bounds are in fact quite rough, they are sufficient to gain qualitative insight into
the limit behavior of the dimensions $d_{n,r}$ when $n$ or $r$ are large. Let us first examine
$\lim_{r\to\infty}d_{n,r}$ for fixed $n$. Denote $\lambda = 4^n-1$. Then, using the
Stirling approximation $\ln (a!)\approx a\ln a - a$, the upper bound reads \be \ln \left( \begin{array}{c} r +\lambda\\
r\end{array}\right) &=& \ln (r+\lambda)! - \ln r! - \ln \lambda!\nonumber\\
&\approx& (r+\lambda)\ln (r+\lambda) - r\ln r - \ln \lambda! - \lambda\nonumber\\&=&  \ln (1 +
\frac{\lambda}{r})^r + \lambda\ln (r+\lambda) - \ln \lambda! - \lambda\nonumber\\ &\approx&
\lambda\ln (r+\lambda) - \ln \lambda!,\ee where in the last line we have used
$(1+\frac{\lambda}{r})^r\approx \exp({\lambda})$ when $r$ is large. Finally, we obtain \be d_{n,r}
\leq \frac{1}{\lambda!} (r + \lambda)^{\lambda}.\ee We have proven:
\begin{thm}
For every fixed $n\in\Bbb{N}_0$, the dimension $d_{n,r}$ tends polynomially in $r$ to infinity . In
other words, for every $n$ there exists a polynomial $p_n(r)$ in $r$ such that $ d_{n,r} = {\cal
O}\left( p_n(r)\right)$.
\end{thm}
Note that a similar result does not hold for $\lim_{n\to \infty} d_{n,r}$ for fixed $r$. Indeed,
the lower bound in lemma 2 shows that \be d_{n,r}\geq {\cal O}\left(\frac{1}{r!} \left(
\frac{4^r}{6}\right)^n\right), \ee which is nonpolynomial in $n$ if $r\geq 2$.

\section{Invariants of degrees 1, 2 and 3}

In this section we investigate the invariants of ${\cal C}_n^l$ of low degrees in more detail. In
particular, we will show the following result:
\begin{thm}
Every invariant of ${\cal C}_n^l$ of degree 1, 2 or 3 is an invariant of $U(2)^{\otimes n}$ (which
also acts by conjugation) and vice versa.
\end{thm}
One of the implications in the theorem is trivial. Indeed, every invariant of $U(2)^{\otimes n}$ is
an invariant of ${\cal C}_n^l$, as the latter is a subgroup of the former. Let us now prove the
reverse implication.

Let $\rho$ be a $2^n\times 2^n$ matrix of variables. Firstly, it follows from theorems 1 and 2 that
${\cal C}_n^l$ has only one invariant of degree 1, namely Tr$(\rho)$, which is trivially an
invariant of $U(2)^{\otimes n}$.

In order to examine the invariants of degrees 2 and 3, it will be convenient to introduce the
following functions:
\begin{defn}
Let $\omega\subseteq\{1, \dots, n\}$. Define the functions $\delta_{\omega}, \epsilon_{\omega}:
\Bbb{F}_2^{2n}\to \Bbb{C}$ by \begin{itemize}
\item[] $\delta_{\omega}(w) = 1$ if supp$(w)=\omega$ and $\delta_{\omega}(w) = 0$ otherwise
\item[] $\epsilon_{\omega}(w) = 1$ if supp$(w)\subseteq\omega$ and $\epsilon_{\omega}(w) = 0$
otherwise.
\end{itemize}
\end{defn}
It is straightforward to show the following relations \be\label{relations1}
&&\epsilon_{\omega} = \sum_{\omega'\subseteq\omega} \delta_{\omega'}\nonumber\\
&&\delta_{\omega}=(-1)^{|\omega|}\sum_{\omega'\subseteq\omega}(-1)^{|\omega'|}\
\epsilon_{\omega'},\ee the first of which is trivial and the second of which can easily be verified
by substitution in the first one.

Now, regarding $r=2$, using example 1 we find that the polynomials \be\label{degree2}
p_{\omega}(\rho)&=&\sum_{w \in \Bbb{F}_2^{2n}, \mbox{\scriptsize\ supp}(w)= \omega } \mbox{Tr }
\left(\sigma_w \otimes \sigma_w \cdot \rho^{\otimes 2}\right)\nonumber\\&=& \sum_{w \in
\Bbb{F}_2^{2n}, \mbox{\scriptsize\ supp}(w)= \omega } \mbox{Tr } \left\{(\sigma_w \cdot
\rho)^2\right\}\ee where $\omega$ ranges over all $2^n$ subsets of $\{1, \dots, n\}$, form a
generating set of the space of invariants of degree 2. Moreover, using the techniques of the
previous section, one can easily show that the $p_{\omega}$'s are linearly independent and
therefore the dimension of this space is $2^n$. Interesting variants of (\ref{degree2}) are the
polynomials \be q_{\omega}(\rho)&=& \sum_{w \in \Bbb{F}_2^{2n}, \mbox{\scriptsize\
supp}(w)\subseteq \omega } \mbox{Tr } \left\{(\sigma_w \cdot \rho)^2\right\}\nonumber\\&=&\mbox{Tr
}\left\{(\mbox{Tr}_{\bar\omega}\ \rho)^2\right\},\ee where the operation Tr$_{\bar\omega}$ denotes
the partial trace over all qubits outside the set $\omega$. The polynomials $q_{\omega}$ are
manifestly invariant under the entire local unitary group. In fact, it is well known that these
polynomials are generators of the space of invariants of $U(2)^{\otimes n}$ of degree two
\cite{rainspol}. Moreover, one has the relations \be\label{relations2}
&&q_{\omega} = \sum_{\omega'\subseteq\omega} p_{\omega'}\nonumber\\
&&p_{\omega}=(-1)^{|\omega|}\sum_{\omega'\subseteq\omega}(-1)^{|\omega'|}\ q_{\omega'},\ee which
follow immediately from (\ref{relations1}). In particular, the second expression in
(\ref{relations2}) shows that every polynomial $p_{\omega}$ is an invariant of $U(2)^{\otimes n}$,
implying that the sets \{$p_{\omega}$\} and $\{q_{\omega}\}$ span the same space, which yields the
desired result for theorem 6 for $r=2$. Furthermore, it follows from (\ref{relations2}) that
polynomials $q_{\omega}$ are a basis as well, being a generating set of cardinality $2^n$ in a
$2^n$-dimensional space.

A similar result can be proven for the invariants of degree 3. Theorem 2 shows that the space of
invariants of ${\cal C}_n^l$ of degree 3 is spanned by all polynomials \be p^{\gamma}_{n,3} =
\sum_{(w^{(1)}, w^{(2)},w^{(3)})\in\gamma} \mbox{ Tr
}(\sigma_{w^{(1)}}\otimes\sigma_{w^{(2)}}\otimes \sigma_{w^{(3)}}\ \rho^{\otimes 3}) \nonumber,\ee
where $\gamma$ ranges over all elements in ${\cal O}_3^n$. Note that, for every $\gamma\in{\cal
O}_3^n$, one has $w^{(1)}+ w^{(2)}+w^{(3)}=0$ whenever $(w^{(1)}, w^{(2)},w^{(3)})\in\gamma$, by
definition of ${\cal O}_3^n$. Using the description of $\gamma$ by sets $\omega^{(i)}$ and
$\omega^{(ij)}$ introduced in theorem 3, it follows that  \be p^{\gamma}_{n,3} = \sum \mbox{ Tr
}(\sigma_{w^{(1)}}\otimes\sigma_{w^{(2)}}\otimes \sigma_{w^{(1)}+ w^{(2)}}\ \rho^{\otimes 3}),\ee
where the sum runs over all couples $(w^{(1)}, w^{(2)})\in
(\Bbb{F}_2^{2n})^{\times 2}$ such that \be &\mbox{supp}(w^{(1)})=\omega_1,\ \mbox{supp}(w^{(2)})=\omega_2&\nonumber\\
 &\mbox{supp}(w^{(1)}+ w^{(2)})=\omega_{12}&,\ee for some $\omega_1, \omega_2,
\omega_{12}\subseteq\{1, \dots, n\}$. Using (\ref{relations1}), a straightforward calculation shows
that $p^{\gamma}_{n,3}$ is, up to an overall sign,  equal to \be\label{degree3} \sum\
(-1)^{|\omega_1'| + |\omega_2'| + |\omega_{12}'|} \mbox{ Tr }\left\{(\mbox{Tr}_{\bar\omega_1'}
\rho)\ (\mbox{Tr}_{\bar\omega_2'} \rho)\ (\mbox{Tr}_{\bar\omega_{12}'} \rho)\right\}\nonumber,\\
\ee where the sum runs over all $\omega_1'\subseteq\omega_1$, $\omega_2'\subseteq\omega_2$ and $
\omega_{12}'\subseteq\omega_{12}$. As the summands in (\ref{degree3}) are manifestly invariant
under the action of $U(2)^{\otimes n}$, the polynomial $p^{\gamma}_{n,3}$ is an invariant of the
local unitary group and the proof of theorem 7 is completed.

\section{Conclusion}

We have performed a systematic study of the invariant algebra of the local Clifford group ${\cal C
}_n^l$, using the description of this group in terms of binary arithmetic. Our approach was to
consider the spaces of homogeneous invariants degree per degree and to construct bases of these
spaces for each degree $r$. In order to study these spaces of homogeneous invariants, we
transformed the problem to the study of certain algebras ${\cal A}_{n,r}$ of matrices, such that
every matrix in an algebra ${\cal A}_{n,r}$ corresponds to an invariant polynomial of degree $r$.
We then constructed bases $\{A_{\gamma}\}_{\gamma\in{\cal O}_r^n}$ of these algebras, which yielded
generating, though linearly dependent, sets $\{p^{\gamma}_{n,r}\}_{\gamma}$ of homogeneous
invariants. We subsequently showed how a basis of invariants could be pinpointed amongst these
polynomials for each degree $r$, which was the main result of this paper.

As stated in the introduction, we believe that these results are relevant in a number of fields in
quantum information theory, with in particular, the classification of binary quantum codes. In
forthcoming work we will apply the present results to this problem.

\begin{acknowledgments}

Dr. Bart De Moor is a full professor at the Katholieke Universiteit Leuven, Belgium. Research
supported by Research Council KUL: GOA-Mefisto 666, GOA-Ambiorics, several PhD/postdoc and fellow
grants; Flemish Government: -   FWO: PhD/postdoc grants, projects, G.0240.99 (multilinear algebra),
G.0407.02 (support vector machines), G.0197.02 (power islands), G.0141.03 (Identification and
cryptography), G.0491.03 (control for intensive care glycemia), G.0120.03 (QIT), G.0452.04 (QC),
G.0499.04 (robust SVM), research communities (ICCoS, ANMMM, MLDM); -   AWI: Bil. Int. Collaboration
Hungary/ Poland; -   IWT: PhD Grants, GBOU (McKnow) Belgian Federal Government: Belgian Federal
Science Policy Office: IUAP V-22 (Dynamical Systems and Control: Computation, Identification and
Modelling, 2002-2006), PODO-II (CP/01/40: TMS and Sustainibility); EU: FP5-Quprodis;  ERNSI; Eureka
2063-IMPACT; Eureka 2419-FliTE; Contract Research/agreements: ISMC/IPCOS, Data4s, TML, Elia, LMS,
IPCOS, Mastercard; QUIPROCONE; QUPRODIS.
\end{acknowledgments}

\bibliographystyle{unsrt}
\bibliography{invar_cliff_PRA}
\end{document}